\documentclass[manuscript,screen]{acmart}

\newlength\mylen
\usepackage{algorithm}
\usepackage{algpseudocode}
\usepackage{tabularx}
\usepackage{tikz}
\usepackage{array}
\newcolumntype{C}{>{\hfil$}p{\mylen}<{$\hfil}}
\usetikzlibrary{matrix}

\newcommand{\bc}[1]{\ensuremath{\mbox{bc}(#1)}}
\newcommand{\cdiv}[1]{\ensuremath{\mbox{cdiv}(#1)}}

\newcommand{\hadj}[2]{\ensuremath{\mbox{hadj}_{#1}(#2)}}

\newcommand{\Tree}[1]{\ensuremath{T_{r}[#1]}}

\makeatletter
\algnewcommand{\LineComment}[1]{\Statex \hskip\ALG@thistlm /*\ #1\ */}
\newcommand{\multiline}[1]{%
	\begin{tabularx}{\dimexpr\linewidth-\ALG@thistlm}[t]{@{}X@{}}
		#1
	\end{tabularx}
}
\makeatother
\algnewcommand{\IIf}[1]{\State\algorithmicif\ #1\ \algorithmicthen}
\algnewcommand{\EndIIf}{\unskip\ \algorithmicend\ \algorithmicif}
\algnewcommand{\FFor}[1]{\State\algorithmicfor\ #1\ \algorithmicdo}
\algnewcommand{\EndFFor}{\unskip\ \algorithmicend\ \algorithmicfor}
\algrenewcommand\algorithmicindent{1.0em}%
\algnewcommand{\IfThenElse}[3]{% \IfThenElse{<if>}{<then>}{<else>}
\State \algorithmicif\ #1\ \algorithmicthen\ #2\ \algorithmicelse\ #3}

\begin{document}

\title{A comparison of two effective methods for reordering columns within supernodes}

\author{M. Ozan Karsavuran}
\affiliation{%
  \institution{Lawrence Berkeley National Laboratory}
  \country{USA}
}
\author{Esmond G. Ng}
\affiliation{%
  \institution{Lawrence Berkeley National Laboratory}
  \country{USA}
}
%\email{webmaster@marysville-ohio.com}
%\affiliation{%
%  \institution{Institute for Clarity in Documentation}
%  \streetaddress{P.O. Box 1212}
%  \country{USA}

\author{Barry W. Peyton}
\affiliation{%
  \institution{Dalton State College}
  \country{USA}
}
%\email{larst@affiliation.org}


%\begin{abstract}
%  A clear and well-documented \LaTeX\ document is presented as an
%  article formatted for publication by ACM in a conference proceedings
%  or journal publication. Based on the ``acmart'' document class, this
%  article presents and explains many of the common variations, as well
%  as many of the formatting elements an author may use in the
%  preparation of the documentation of their work.
%\end{abstract}
\begin{abstract}
In some recent papers,
researchers have found two very good methods
for reordering columns within supernodes in sparse Cholesky factors;
these reorderings can be very useful for certain factorization methods.
The first of these reordering methods is based on modeling the underlying
problem as a {\em traveling salesman problem}~(TSP),
and the second of these methods is based on {\em partition refinement}~(PR).
In this paper, we devise a fair way to compare the two methods.
While the two methods are virtually the same in the quality
of the reorderings that they produce,~PR should be the method
of choice because~PR reorderings can be computed using far less time
and storage than~TSP reorderings.
\end{abstract}


%\begin{CCSXML}
%<ccs2012>
%  <concept_id>00000000.0000000.0000000</concept_id>
%  <concept_desc>Do Not Use This Code, Generate the Correct Terms for Your Paper</concept_desc>
%  <concept_significance>500</concept_significance>
%  <concept_id>00000000.00000000.00000000</concept_id>
%  <concept_desc>Do Not Use This Code, Generate the Correct Terms for Your Paper</concept_desc>
%  <concept_significance>300</concept_significance>
%  <concept_id>00000000.00000000.00000000</concept_id>
%  <concept_desc>Do Not Use This Code, Generate the Correct Terms for Your Paper</concept_desc>
%  <concept_significance>100</concept_significance>
%  <concept_id>00000000.00000000.00000000</concept_id>
%  <concept_desc>Do Not Use This Code, Generate the Correct Terms for Your Paper</concept_desc>
%  <concept_significance>100</concept_significance>
%</ccs2012>
%\end{CCSXML}

%\ccsdesc[500]{Do Not Use This Code~Generate the Correct Terms for Your Paper}
%\ccsdesc[300]{Do Not Use This Code~Generate the Correct Terms for Your Paper}
%\ccsdesc{Do Not Use This Code~Generate the Correct Terms for Your Paper}
%\ccsdesc[100]{Do Not Use This Code~Generate the Correct Terms for Your Paper}



\maketitle

%
% partitioning into paths for Suporder.tex
%

\section{Introduction}
\label{sec:intro}

Let~$A$ be an $n$~by~$n$ sparse symmetric positive definite matrix, and
let $A=LL^T$ be the Cholesky factorization of~$A$,
where~$L$ is a lower triangular matrix.
It is well known that~$L$ suffers {\em fill} during such a factorization;
that is, $L$ will have nonzero entries in locations occupied
by zeros in~$A$.
As a practical matter, it is important to limit the number of such fill entries
in~$L$.
Consequently, software for solving a sparse symmetric positive definite
linear system $Ax=b$ via sparse Cholesky factorization requires
the following four steps.

First,
compute a fill-reducing ordering of~$A$ using either the 
{\em nested dissection}~\cite{George73,KK99}
or the
{\em minimum degree}~\cite{ADD96,GL89,Liu-MMD-1985,Tinney-Walker67}
ordering heuristic (the {\bf ordering} step).
Second, compute the needed information concerning and data structures for
the sparse Cholesky factor matrix (the {\bf symbolic factorization} step).
Third, compute the sparse Cholesky factor within the data structures
computed during the symbolic factorization step (the {\bf numerical factorization} step).
Fourth, solve the linear system by performing in succession
a sparse forward solve and a sparse backward solve using
the sparse Cholesky factor and its transpose, respectively
(the {\bf solve} step).
It will be convenient henceforth to assume that the matrix~$A$
has been reordered by the fill-reducing ordering obtained during the first step,
so that $A=LL^T$.

The authors of this paper, along with J.~L.~Peyton, presented a thorough
look~\cite{KNPP24} at some serial algorithms for the third step in the
solution process (the numerical factorization step).
Our goal was to improve the performance of serial sparse Cholesky
factorization algorithms on multicore processors when only the
multithreaded BLAS are used to parallelize the computation.
Essentially, our first paper~\cite{KNPP24} explored what can be done for serial 
sparse Cholesky factorization using the techniques and methodology
used in LAPACK.

Our primary contribution in~\cite{KNPP24} is the factorization method
that we called {\em right-looking blocked} (RLB).
%(The reader should consult~\cite{KNPP24} for a complete description of~RLB.)
Like all of the other factorization methods studied in~\cite{KNPP24},
RLB relies on {\em supernodes} to obtain efficiency,
where {\em supernodes} are, roughly speaking, sets of consecutive
columns in the factor matrix sharing the same zero-nonzero structure.
RLB, however, is unique among the factorization methods studied in~\cite{KNPP24}
in that it requires no floating-point working storage or assembly
operations.
RLB is also unique among the factorization methods studied in~\cite{KNPP24}
in that it is {\em entirely} dependent for efficiency on the existence
of few and large dense blocks joining together pairs of supernodes
in the factor matrix.
Furthermore, the number of and size of these dense blocks are {\em crucially}
dependent on how the columns of the factor matrix are ordered
{\em within} supernodes.
As a result, RLB is perfectly suited for studying the {\em quality}
of algorithms for reordering columns within supernodes.
It is precisely a study of this sort that will occupy our attention
in this paper.

Our paper is organized as follows.
In Section~\ref{sec:example}, we explore the application of~RLB
to a small example, highlighting the importance of reordering
columns within supernodes to the efficiency of method~RLB.
To make the paper self-contained, we give a high-level
description of~RLB in Section~\ref{sec:RLBalg}.
There are two known effective methods for reordering
columns within supernodes---one based on the {\em traveling salesman problem}~\cite{PFRR17} (TSP),
and the other based on {\em partition refinement}~\cite{JNP18} (PR).
Section~\ref{sec:reorder} both gives the proper references for~TSP and~PR, and it
sketches out how~TSP and~PR will both be improved in this paper before they
are ultimately compared with each other in as fair a manner as possible.
In Section~\ref{sec:tests}, we present a description of our test bed;
in the main body of the section,
we present improvements to both~TSP and~PR, and
we present timing results and storage statistics that
confirm the superiority of~PR.
Section~\ref{sec:conclusion} presents our concluding remarks.

%We will spend most of the rest of the introduction delving into an example
%used in~\cite{KNPP24}, but we use it to illustrate what we have been talking about thus far;
%we delve into this example in greater detail here, since we have
%more to explain here than we did in~\cite{KNPP24}.

\section{Method RLB}
\label{sec:RLB}

\subsection{Method RLB applied to an example}
\label{sec:example}
We first introduce RLB by sketching out its application
to the small sparse Cholesky factor shown in Figure~\ref{fig:supernode1}.
Before doing so, we will need the following notation as we examine the figure;
for a matrix~$C$ and two index sets~$K$ and~$J$ we will let
$C_{K,J}$ be the submatrix of~$C$ with rows taken from~$K$
and columns taken from~$J$.

\begin{figure}[htp]
\footnotesize
\setlength{\arraycolsep}{3pt}
\begin{center}
      %$L =
      %$\begin{array}{c}
      \begin{eqnarray*}
      {\color{white} L} & {\color{white} =} &
      {\color{white} \left[
       \begin{array}{cc|cc|ccccc}
         \multicolumn{2}{c}{\color{black} J_1} &
         \multicolumn{2}{c}{\color{black} J_2} &
         \multicolumn{5}{c}{\color{black} J_3} \\
           {\color{white} 1} & 
           {\color{white} +} & 
           {\color{white} 3} & 
           {\color{white} +} & 
           {\color{white} +} & 
           {\color{white} +} & 
           {\color{white} +} & 
           {\color{white} 8} & 
           {\color{white} 9}
       \end{array}
       \right]} \\
       L & = &
       \left[
       \begin{array}{cc|cc|ccccc}
         {1} & & \multicolumn{7}{c}{} \\
         \ast & {2} & \multicolumn{7}{c}{} \\
         & & 3 & & \multicolumn{5}{c}{} \\
         & & \ast & 4 & \multicolumn{5}{c}{} \\
         \ast & \ast & \ast & \ast & 5 & \multicolumn{4}{c}{} \\
         \ast & + & & & \ast & 6 & \multicolumn{3}{c}{} \\
         & & \ast & + & + & \ast & 7 & \multicolumn{2}{c}{} \\
         & & \ast & \ast & \ast & + & \ast & 8 & \multicolumn{1}{c}{} \\
         \ast & \ast & & & + & \ast & + & \ast & 9
       \end{array}
       \right]
      \end{eqnarray*}
      %\end{array}$
  \caption{The supernodes of a sparse Cholesky factor $L$.  
	Each symbol~`$\ast$' signifies
	an off-diagonal entry that is nonzero in both~$A$ and~$L$; 
	each symbol~`$+$' signifies an off-diagonal entry that is zero in~$A$ but nonzero
	in~$L$---a fill entry in~$L$.}
  \label{fig:supernode1}
\end{center}
\end{figure}

In the figure, the first supernode $J_{1} = \{1,2\}$ comprises columns $1$ and $2$.
The $2$~by~$2$ submatrix $L_{J_{1},J_{1}}$ is a dense lower triangular matrix.
Below this dense lower triangular block, columns $1$ and $2$ of $L$
share the same sparsity structure;
specifically, both have nonzeros in rows $5$,~$6$, and~$9$ only.
The second supernode $J_{2} = \{3,4\}$ comprises columns $3$ and $4$.
The $2$~by~$2$ submatrix $L_{J_{2},J_{2}}$ is a dense lower triangular matrix.
Below this dense lower triangular block, columns $3$ and $4$ of $L$
share the same sparsity structure;
specifically, both have nonzeros in rows $5$,~$7$, and~$8$ only.
The third supernode $J_{3} = \{5,6,7,8,9\}$ comprises columns $5$ through $9$.
The $5$~by~$5$ submatrix $L_{J_{3},J_{3}}$ is a dense lower triangular matrix.

To compute~$L$, the algorithm RLB first computes supernode~$J_{1}$, 
which requires no updates from the left.
This is done by first calling DPOTRF to factorize~$L_{J_1,J_1}$, 
followed by calling DTRSM (dense block triangular solve) 
to compute the rectangular part of the supernode.
Supernode~$J_{1}$ has two dense {\em blocks} joining it to 
supernode~$J_{3}$, namely, 
$B=\{5,6\}$ and $B'=\{9\}$,
where each set identifies the rows in the block.
The block~$B$ contains~$5$ and~$6$ because both $L_{5,J_{1}}$ and $L_{6,J_{1}}$
are dense rows;
the block~$B$ does not contain either $L_{4,J_{1}}$ or $L_{7,J_{1}}$ because
each is a zero row.
($L_{4,J_1}$ cannot belong to~$B$ in any case, because we have
$4 \in J_2$, not $4 \in J_3$.)
The block~$B'$ contains~$9$ because $L_{9,J_{1}}$ is a dense row;
the block~$B'$ does not contain $L_{8,J_{1}}$ because it is a zero row,
and, clearly, $B'$ can contain no row below $L_{9,J_{1}}$.
The $(J_3,J_1)$-blocks are precisely~$B$ and~$B'$,
and the {\em block count} $\bc{J_3,J_1}$ = 2.

At this point, RLB now updates supernode~$J_{3}$ with supernode~$J_{1}$,
as follows.
It first computes the lower triangle of the following:
\[
  L_{B,B} \gets L_{B,B} - L_{B,J_{1}} * L_{B,J_{1}}^T
\]
using the BLAS routine DSYRK.
It then computes
\[
  L_{B',B} \gets L_{B',B} - L_{B',J_{1}} * L_{B,J_{1}}^T
\]
using the BLAS routine DGEMM.
It then computes the lower triangle of the following:
\[
  L_{B',B'} \gets L_{B',B'} - L_{B',J_{1}} * L_{B',J_{1}}^T
\]
using the BLAS routine DSYRK.
Since $L_{J_2,J_1}=0$,
there are no updates from supernode~$J_{1}$ to supernode~$J_{2}$.

After supernode~$J_{1}$ is processed,
RLB will complete supernode~$J_{2}$.
Supernode~$J_{2}$ has two dense {\em blocks} joining it to 
supernode~$J_{3}$, namely, 
$B=\{5\}$ and $B'=\{7,8\}$.
The $(J_3,J_2)$-blocks are precisely~$B$ and~$B'$, as given above,
and the block count $\bc{J_3,J_2}$ = 2.
After supernode~$J_2$ has been completed,
supernode~$J_{2}$ will update supernode~$J_{3}$
with these blocks in the same way that supernode~$J_{1}$ did with
its blocks.
After that,
supernode~$J_{3}$ has received all of its updates,
and RLB finishes the computation by completing supernode~$J_{3}$.

Intuitively, the performance of~RLB would improve if the blocks could
somehow be made fewer and larger, but without increasing fill.
It turns out that we can achieve that goal by reordering the columns
within supernodes.

Let us turn our attention again to the sparsity pattern of the
Cholesky factor~$L$ shown in Figure~\ref{fig:supernode1}.
Recall that the supernodes were identified as
$J_{1} = \{1,2\}$,
$J_{2} = \{3,4\}$, and
$J_{3} = \{5,6,7,8,9\}$.
Let us now symmetrically permute the rows and columns of supernode $J_{3}$.
Specifically,
let us move row/column 6 to row/column 5 (i.e., $6 \rightarrow 5$),
along with $9 \rightarrow 6$,
$5 \rightarrow 7$,
$7 \rightarrow 8$, and
$8 \rightarrow 9$.
The sparsity pattern of the new Cholesky factor~$\widehat{L}$
is shown in Figure~\ref{fig:supernode2}.
\begin{figure}[htp]
\footnotesize
\setlength{\arraycolsep}{3pt}
\begin{center}
 %$\hat{L}  =  
 %$\begin{array}{c}
  \begin{eqnarray*}
    {\color{white} \widehat{L}} & {\color{white} =} &
    {\color{white} \left[
    \begin{array}{cc|cc|ccccc}
      \multicolumn{2}{c}{\color{black} J_1} &
      \multicolumn{2}{c}{\color{black} J_2} &
      \multicolumn{5}{c}{\color{black} J_3} \\
%     {\color{white} 1} & 
%     {\color{white} +} & 
%     {\color{white} 3} & 
%     {\color{white} +} & 
%     {\color{white} +} & 
%     {\color{white} 6} & 
%     {\color{white} +} & 
%     {\color{white} +} & 
%     {\color{white} 9}
      {\color{white} 1} & 
      {\color{white} +} & 
      {\color{white} 3} & 
      {\color{white} +} & 
      {\color{black} 5} & 
      {\color{black} 6} & 
      {\color{black} 7} & 
      {\color{black} 8} & 
      {\color{black} 9}
    \end{array}
    \right]} \\
    \widehat{L} & = &
    \left[
      \begin{array}{cc|cc|ccccc}
        $1$ & & \multicolumn{7}{c}{} \\
        \ast & $2$ & \multicolumn{7}{c}{} \\
        & & $3$ & & \multicolumn{5}{c}{} \\
        & & \ast & $4$ & \multicolumn{5}{c}{} \\ 
        \ast & + & & & {6} & \multicolumn{4}{c}{} \\
        \ast & \ast & & & \ast & {9} & \multicolumn{3}{c}{} \\ 
        \ast & \ast & \ast & \ast & \ast & + & {5} & \multicolumn{2}{c}{} \\
        & & \ast & + & \ast & + & + & {7} & \multicolumn{1}{c}{} \\
        & & \ast & \ast & + & \ast & \ast & \ast & {8} \\
      \end{array}
    \right]
  \end{eqnarray*}
 %\end{array}$
  \caption{The supernodes of the sparse Cholesky factor $\widehat{L}$
  obtained after a symmetric permutation of supernode~$J_3$ in Figure~\ref{fig:supernode1}.
  Let $\widehat{A}$ be the new version of $A$ after the symmetric permutation.
  Each symbol~`$\ast$' signifies
  an off-diagonal entry that is nonzero in both~$\widehat{A}$ and~$\widehat{L}$; 
  each symbol~`$+$' signifies an off-diagonal entry that is zero in~$\widehat{A}$
  but nonzero in~$\widehat{L}$.}
  \label{fig:supernode2}
\end{center}
\mbox{}
\end{figure}
(The old numbers for supernode~$J_{3}$ are shown on the main diagonal,
and the new numbers for supernode~$J_{3}$ are shown above its columns.)
Such a reordering does not change the nonzero count of~$L$.

Consider now what happens when~RLB is used to compute~$\widehat{L}$.
Supernode~$J_{1}$ now has only one dense block $B=\{5,6,7\}$ joining it to
supernode~$J_{3}$.
The $(J_3,J_1)$-block is precisely~$B$,
and the block count $\bc{J_3,J_1}$ = 1.
After supernode~$J_1$ has been completed,
RLB then updates supernode~$J_{3}$ with supernode~$J_{1}$,
as follows.
It computes the lower triangle of the following:
\[
  L_{B,B} \gets L_{B,B} - L_{B,J_{1}} * L_{B,J_{1}}^T
\]
using the BLAS routine DSYRK.
We see that there is
now  only one block update from supernode~$J_{1}$
to supernode~$J_{3}$, rather than the three that were required under
the original ordering.

Supernode~$J_{2}$ now also has only one dense block $B'=\{7,8,9\}$ joining it to
supernode~$J_{3}$.
The $(J_3,J_2)$-block is precisely~$B'$,
and the block count $\bc{J_3,J_2}$ = 1.
After supernode~$J_2$ has been completed,
RLB then updates supernode~$J_{3}$ with supernode~$J_{2}$,
as follows.
It computes the lower triangle of the following:
\[
  L_{B',B'} \gets L_{B',B'} - L_{B',J_{2}} * L_{B',J_{2}}^T
\]
using the BLAS routine DSYRK.
We see that there is
now  only one block update from supernode~$J_{2}$
to supernode~$J_{3}$, rather than the three that were required under
the original ordering.

Recent research has led to excellent algorithms for reordering
columns within supernodes to reduce the number of blocks in this fashion.
Our discussion of the example and the notation introduced there
suggest the key underlying combinatorial optimization problem.
Let~$J_k$ be the~$k^{\rm th}$ supernode in a sparse Cholesky factor.
We would like to reorder the columns within supernode~$J_k$ so that
\begin{equation} \label{eqn:supord}
  \sum_{r=1}^{k-1} \bc{J_k,J_r}
\end{equation}
is minimized.
The reader may easily verify that for the Cholesky factor 
in Figure~\ref{fig:supernode2},
\begin{displaymath}
  \sum_{r=1}^{2} \bc{J_3,J_r} = 2
\end{displaymath}
is the minimum possible over all reorderings of~$J_3$.

\subsection{A high-level description of method RLB}
\label{sec:RLBalg}

To make this paper adequately self-contained,
we include a high-level description of {\em right-looking blocked}
sparse Cholesky~(RLB) in Algorithm~\ref{alg:RLB}.
\begin{algorithm}[htp] % :set ts=2
%\footnotesize
%\small
\begin{flushleft}
	{\bf Input:}
	The supernodal elimination tree parent function $p(J)$ for every supernode~$J$; \\
	the set of blocks for each supernode $J$; \\
	the data structure for~$L$ initialized with the appropriate entries of~$A$. \\
	{\bf Output:}
	$L_{*,J}$, for every supernode~$J$.
\end{flushleft}
				\caption{A high-level description of our {\em right-looking blocked} sparse Cholesky algorithm (RLB).}
\label{alg:RLB}
	\begin{algorithmic}[1]
	\For{each supernode $J$ (in ascending order)}
		\State{Perform $\cdiv{J}$ (DPOTRF, DTRSM);}
		\LineComment{Update $J$'s ancestors.}
		\State{$P \gets p(J)$;}
		\While{there remains an ancestor supernode of $J$ to update}
			\For{each block $B$ of $J$ {\bf such that} $B \subseteq P$ (in order)}
				\State \multiline{%
					Perform the lower triangle of the update
					$L_{B,B} \gets L_{B,B} - L_{B,J} * L_{B,J}^T$ (DSYRK);}
					\For{each ``maximal'' block $B'$ of $J$ below $B$ (in order)}
						\State \multiline{%
							Perform the update
							$L_{B',B} \gets L_{B',B} - L_{B',J} * L_{B,J}^T$ (DGEMM);}
					\EndFor
			\EndFor
			\State{$C \gets P$; $P \gets p(P)$;}
			\Comment{Next ancestor $P$}
		\EndWhile
	\EndFor
\end{algorithmic}
\end{algorithm}
Input into the algorithm are the {\em supernodal elimination tree}'s
parent function and the set of blocks for each supernode~$J$.
(For background on elimination trees,
see Liu~\cite{Liu90}.)
The data structure that will eventually contain~$L$ initially
contains the appropriate entries of~$A$.
On output, the data structure for~$L$ will contain~$L_{*,J}$,
for every supernode~$J$.

The outer loop from line~1 to line~13 processes every supernode~$J$
in ascending order.
Before line~2 is executed, supernode~$J$ has received all of its updates
from the appropriate supernodes;
these updates were received during prior iterations through
the {\bf while} loop from line~4 to line~12 when the outer
{\bf for} loop was processing supernodes~$J'$ that precede~$J$.
The $\cdiv{J}$ operation in line~2 first uses dense Cholesky
factorization (DPOTRF) to compute~$L_{J,J}$,
and then uses dense triangular solution (DTRSM) to compute
the columns of~$L_{*,J}$ below~$L_{J,J}$.

The algorithm then uses~$L_{*,J}$ to update~$J$'s ancestors in the
supernodal elimination tree.
In line~3,~$P$, the parent of~$J$, becomes the first ancestor of~$J$
to visit.
In lines~5--6, for each block~$B$ of~$J$ such that $B \subseteq P$,
first perform the lower triangle of the update
$L_{B,B} \gets L_{B,B} - L_{B,J} * L_{B,J}^T$ (DSYRK),
which takes care of the diagonal block update
using block~$B$.
Subsequently, in lines~7--9,
for each ``maximal'' block~$B'$ of~$J$ {\em below}~$B$,
we perform the update
$L_{B',B} \gets L_{B',B} - L_{B',J} * L_{B,J}^T$ (DGEMM),
which takes care of the updates below the diagonal block using block~$B$.
(A ``maximal'' block may consist of two contiguous blocks joining
a supernode to two consecutive supernodes.)
The {\bf while} loop from line~4 to line~12 completes all of the 
updates from supernode~$J$ to its ancestor supernodes in this
fashion.
We emphasize that the updates are performed, in place,
directly from and to factor storage;
that is, the algorithm uses no floating-point working
storage or assembly operations.

The reader should consult
Karsavuran, Ng, Peyton, and Peyton~\cite{KNPP24}
for a more detailed description of~RLB.
In particular, the data structures in use are spelled out there
quite clearly.
Also, Karsavuran, et al.~\cite{KNPP24} gives the details
of the indexing techniques used to handle sparsity issues
by both right-looking algorithms introduced in that paper.

\section{An overview of the contents of this paper}
\label{sec:reorder}
Pichon, Faverge, Ramet, and Roman~\cite{PFRR17}
were the first to take seriously the problem
of reordering columns within supernodes, in that they were the first to
treat it in a highly technical manner.
They ingeniously 
formulated the underlying optimization problem
given at the end of Section~\ref{sec:example}
%in the sentence containing the expression in~(\ref{eqn:supord})
as a {\em traveling salesman problem}, for which there exist powerful
and effective heuristics.
We will refer to their approach as~TSP.
The problem with their approach was not ordering quality;
it was the cost, in time, of computing the needed~TSP distances~\cite{JNP18,PFRR17}.
In~2021, Jacquelin, Ng, and Peyton~\cite{JNP21} devised a much faster way to 
compute the needed distances, which greatly reduces the runtimes
for the~TSP method.

In~2017,
Jacquelin, Ng, and Peyton~\cite{JNP18} proposed a simpler heuristic
for reordering columns within supernodes based on {\em partition refinement}~\cite{PT87}.
In their paper, they report faster runtimes for their method than~TSP,
while obtaining similar ordering quality.
We will refer to their method as~PR.

We will not describe method~TSP in any detail in this paper, nor will
we describe how to model the underlying combinatorial optimization problem
as a traveling salesman problem.  
For full details, the reader should consult Pichon, et al.~\cite{PFRR17}
and Jacquelin, et al.~\cite{JNP21}.
We will have occasion to describe method~PR, but only at a high level;
hence, the reader may need to consult Jacquelin, Ng, and Peyton~\cite{JNP18}
for full details.

In this paper, we perform a careful comparison of~TSP and~PR;
we compare them, primarily, by measuring the impact of~TSP and~PR
on~RLB factorization times using Intel's MKL multithreaded BLAS
on~48 cores of our test machine. 
(We use here the same experimental test bed that we employed
in our first paper Karsavuran, et al.~\cite{KNPP24},
but the experiments are limited to factorization method~RLB only.)

We introduce two techniques for improving the quality of the~TSP
reorderings;
we observe that the best results for~TSP are obtained when the two
techniques are combined.
We introduce a new way to reorganize the~PR reordering algorithm to
make it much more time and storage efficient.
We also introduce a single technique for modestly improving the quality of the~PR
reorderings.
From that point forward,~TSP refers to the best version of~TSP found above,
and~PR refers to the best version of~PR found above.
When we compare these~TSP reorderings against these~PR reorderings,
two things become abundantly clear:
\begin{enumerate}
  \item
    The quality of the~TSP reorderings and the~PR reorderings
    track each other very closely, and
  \item
    the overhead in both time and storage for the~TSP reorderings is 
    exorbitant relative to that required by the~PR reorderings.
\end{enumerate}
We feel that our experiment here, combined with the level of care and 
development that we have put into the implementation of both methods,
permit an unambiguous conclusion that~PR should be the method of 
choice over~TSP.

%
% partitioning into paths for Suporder.tex
%

\section{Some experiments involving the TSP and PR methods}
\label{sec:tests}

All of the experiments in this paper are described and reported on
in this section.
The section is organized as follows.
Section~\ref{sec:proceedure} details how the testing was carried out.
Section~\ref{sec:TSP} introduces two new features that improve ordering
quality for~TSP,
and it reports on the results obtained using these two new features.
Section~\ref{sec:PR} introduces a new reorganized version of the~PR
reordering algorithm that reduces both runtimes and the working
storage requirements.
It also introduces a new feature that improves ordering
quality for~PR,
and it reports on the results obtained using this new feature.
In Section~\ref{sec:TSP-PR}, we restrict our attention to
the best versions of~TSP and~PR created in Sections~\ref{sec:TSP}
and~\ref{sec:PR}.
Our results in Section~\ref{sec:TSP-PR} confirm that~TSP and~PR
generate orderings of virtually equal quality,
though it appears that~PR may have a very slight advantage over~TSP.
Our results in Section~\ref{sec:TSP-PR} also unambiguously demonstrate
the exorbitant overhead time and workspace required by~TSP
relative to those required by~PR.

%
% partitioning into paths for Suporder.tex
%

\subsection{How the testing was carried out}
\label{sec:proceedure}

In this subsection, we repeat much that can be found
in the analogous subsection of the first paper~\cite{KNPP24},
since the test bed here is nearly the same as that
used in~\cite{KNPP24}, and we wish to make it possible
to read this paper independently of that paper.

For our testing, we selected a set of matrices from the
SuiteSparse collection~\cite{DH11}
of sparse matrices.
We included every symmetric matrix for which $n \geq$ 500,000,
with the following restrictions.
We included only matrices that could be in a realistic sparse
linear system to solve.
Specifically, we excluded graphs from social networks and all other
graphs in no way connected to a linear system.
There is one inconsistency in our selection criteria,
due to practical considerations.
We included the matrices nlpkkt80 and nlpkkt120 (two related optimization matrices),
but we excluded the other matrices in this particular family of matrices
because they require too much storage and time to factor.
Ultimately, there are initially~36 matrices eligible for use in our testing.
(These are precisely the matrices that were used for testing in
our first paper~\cite{KNPP24}.)

In~\cite{KNPP24}, however,~15 of the~36 matrices led to serial factorization
times less than ten seconds;
these were called our {\em small} matrices.
In this paper, there is no need to include these so-called
small matrices; here, we will use exclusively the other~21 matrices,
which were called {\em large} matrices in~\cite{KNPP24}.

In the ordering step of the solution process, 
we use the nested dissection routine from {\sc metis}~\cite{KK99}
to generate the fill-reducing ordering.
This is followed by the symbolic factorization step;
we need to discuss in some detail two tasks that are
performed during this step.

First, we need to discuss how supernodes are produced.
Within the symbolic factorization step,
the {\em fundamental supernode partition}~\cite{LNP93} is computed first.
Most often, however, this supernode partition will be 
so fine near the bottom of the {\em supernodal elimination tree} that it hinders
good factorization performance.
Ashcraft and Grimes~\cite{AG89} introduced the idea of 
merging supernodes together,
thereby coarsening the supernode partition in order
to improve factorization performance.
This has become a standard practice in software for
sparse symmetric factorization.
For example, both the MA87 package~\cite{HRS10}
and the MA57 package~\cite{Duff04} perform such supernode merging by default.
In our symbolic factorization step, we will do likewise;
the details follow.

Our supernode-merging algorithm merges a sequence of child-parent
pairs~$J$ and~$p(J)$ (in the supernodal elimination tree)
until a stopping criterion is satisfied.
As the next pair~$J$ and~$p(J)$ to merge into a single supernode,
the algorithm chooses a pair whose merging creates the minimum
amount of new fill in the factor matrix.
(To do this requires the use of a heap.)
The merging stops whenever the next merging operation will cause
the cumulative percentage increase in factor storage to 
exceed~$12.5$ percent.
(We have tested various values for the percentage,
and~$12.5$ percent works well.)
For every test matrix, the associated increase in factorization 
work never exceeds one percent.
By doing it in this way,
we are trying to normalize supernode merging across
our set of test matrices.

During the symbolic factorization step,
after the supernode partition has been coarsened,
we reorder the columns within supernodes using either a version of the
{\em partition refinement} (PR) method
or a version of the {\em traveling salesman problem} (TSP) method.
Recall that it is
essential to do so for the factorization
method~RLB,
since RLB is not at all a viable method when there is no
reordering of columns within supernodes.
On the other hand, the sensitivity of RLB's performance to the
order of the columns within the supernodes makes RLB's performance
ideal for evaluating the quality of~TSP and~PR reorderings.

We ran the experiments on a dual socket machine containing two Cascade Lake 
Intel(R) Xeon(R) Gold~5220R processors (2.20GHz cpus) 
with~24 cores per socket (48~cores total) and 256~GB of memory.
The 256~GB of memory is split evenly across two NUMA domains, with one NUMA domain per socket.
We compiled our Fortran code with the gfortran compiler
using the optimization flag~-O3.
For all of the runs,
Intel's MKL multithreaded BLAS were linked in.
We ran the experiments with OpenMP affinity enabled
because this improved performance somewhat.
All factorizations were performed using all~48 cores.
Because the timings of the factorizations were somewhat unstable,
we used as our timing, in each case, the median of the timings
of seven consecutive factorizations.

%
% partitioning into paths for Suporder.tex
%

\subsection{Two improvements to the quality of TSP reorderings}
\label{sec:TSP}

In this subsection, we introduce two changes
that improve the quality of~TSP reorderings: one to the~TSP method 
itself and the other to its implementation.
We will show the impact of these changes at the end of the subsection.

The first change is a change in the heuristic used to solve
the actual traveling salesman problem.
Several different heuristics~\cite{JM97} can be used for this purpose;
among these are {\em insertion methods}, which can be described
as follows.
Before the next insertion step, we have a partially completed 
circuit and a set of columns~$J'$ not yet in the circuit.
At the next step, a column in~$J'$ is selected and inserted into
the current circuit at a location that {\em minimizes} the increase
in the length of the modified circuit.
The three commonly used insertion heuristics vary in how the new column
from~$J'$ is chosen.
If it is chosen arbitrarily, then we will refer to this
as the {\em arbitrary insertion} method.
If it is chosen to be a member of~$J'$ that {\em minimizes}
the distance from the new column to the current circuit, then this
is known as the {\em nearest insertion} method.
If it is chosen to be a member of~$J'$ that {\em maximizes}
the distance from the new column to the current circuit, then this
is known as the {\em farthest insertion} method.

Rosenkrantz, Stearns, and Lewis~\cite{RSL77} published a seminal paper on 
insertion methods.
Rosenkrantz, et al.~\cite{RSL77} were able to show that nearest insertion
produces a circuit guaranteed to be within a factor of two of 
optimal in length.
They~\cite{RSL77} were not able to prove any such optimality bounds
for arbitrary insertion or farthest insertion.
Rosenkrantz, et al.~\cite{RSL77} made some allusions to the superior 
performance, in practice, of farthest insertion in restricting circuit lengths,
despite its lack of theoretical underpinnings.
Johnson and McGeoch~\cite{JM97} placed even more emphasis
on the superior performance of farthest insertion in this regard.

We now turn our attention to how solutions to the traveling
salesman problem have been approximated in the~TSP reordering codes
in the literature.
Pichon, et al.~\cite{PFRR17} stated in their paper that they used
nearest insertion.
It turned out that this was incorrect;
they actually used arbitrary insertion in their code.
Jacquelin, et al.~\cite{JNP21} were completely preoccupied
with maintaining a fair comparison with the code in 
Pichon, et al.~\cite{PFRR17}.
So, naturally they used arbitrary insertion simply to match
what was done in Pichon, et al.~\cite{PFRR17}.
Nonetheless, Jacquelin, et al.~\cite{JNP21} built into their~TSP
reordering code the option of using any of the three 
insertion methods: arbitrary insertion, nearest insertion,
and farthest insertion.

In recent experiments, we have found that farthest insertion
results in better reorderings for method~TSP.
We report on these experiments below in this section.
The second improvement to the~TSP reorderings actually
involves a change to the underlying combinatorial
optimization problem.

Let~$J_k$ be the~$k^{\rm th}$ supernode in a sparse Cholesky factor.
As we said in Section~\ref{sec:intro}, the underlying
combinatorial optimization problem is to reorder
supernode~$J_k$ so that
\begin{displaymath}
  \sum_{r=1}^{k-1} \bc{J_k,J_r}
\end{displaymath}
is minimized.
Let~$B$ and~$B'$ be two blocks in supernode~$J_r$,
where~$B$ is a $(J_k,J_r)$-block and
$B'$ is a $(J_{\ell},J_r)$-block for some~$\ell > k$.
RLB will perform the following update to supernode~$J_k$
while it is processing supernode~$J_r$:
\begin{displaymath}
  L_{B',B} \gets L_{B',B} - L_{B',J_r} * L_{B,J_r}^T.
\end{displaymath}
Note that the amount of work required by the matrix-matrix
product is proportional to~$|J_r|$, the width of supernode~$J_r$.
Intuitively, creating large blocks within the wider supernodes
will be more effective than creating large blocks within
the narrower supernodes.

Based on this intuition,
we changed the underlying combinatorial optimization problem to 
the following:
Reorder supernode~$J_k$ so that 
\begin{equation} \label{eqnwts}
	\sum_{r=1}^{k-1} |J_r| \cdot \bc{J_k,J_r}
\end{equation}
is minimized.
We will refer to the new TSP reordering method arising from this
change as the {\em weighted} version.
The needed change to the~TSP reordering code
is trivial and creates no increase in~TSP times.
We do not document it here,
but all that is required are some slight modifications
to some computations involving the 
supernodal elimination tree.

\begin{figure}[htb]
\begin{center}
  \caption{Performance profile for RLB factorization times using four different versions
           of~TSP reorderings.}
  \includegraphics[width=.8\textwidth]{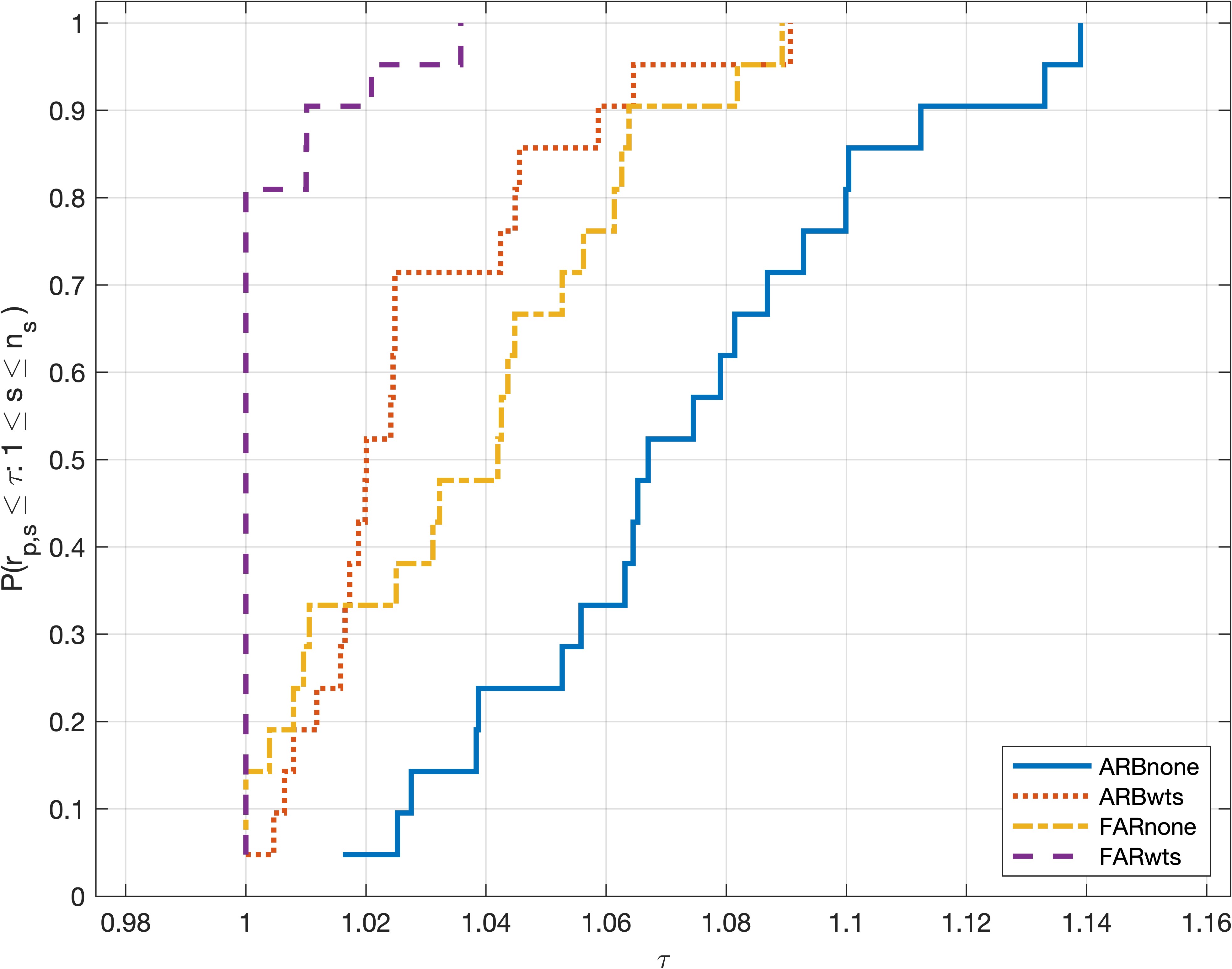}
  \label{fig:TSP}
\end{center}
\end{figure}

In Figure~\ref{fig:TSP}, we display a performance profile 
for~RLB factorization times using
the versions of~TSP
relevant to our discussion in this subsection.
Performing poorest is the ``default'' version of~TSP,
using arbitrary insertion and no weights in the objective function (ARBnone).
This represents the performance of the~TSP algorithms used in Pichon, et al.~\cite{PFRR17}
and Jacquelin, et al.~\cite{JNP21}.
Performing better than ARBnone,
but not best, are the two runs that used only a single enhancement:
one using arbitrary insertion and weights in the objective function
(ARBwts),
and the other using farthest insertion and no weights in 
the objective function (FARnone).

Performing best by far is the version using both farthest insertion
and weights in the objective function (FARwts).
FARwts is the fastest method for approximately eighty percent of the
matrices.
Note also that for three of the four other matrices,
it costs slightly more than two percent 
extra time, or less.
Clearly, farthest insertion and the use of weights in 
objective function~(\ref{eqnwts})
are worthwhile changes to make to the~TSP method.

%
% partitioning into paths for Suporder.tex
%

\subsection{Two improvements to the PR method}
\label{sec:PR}

In this subsection, we introduce two improvements to the~PR reordering method.
First, and most importantly, we reorganize the computation so that one supernode
is partitioned and refined at a time.
This leads to great reductions in both the time and the working storage
required by the computation.
Second, we modestly improve the quality of the~PR reorderings by modifying the
order in which the sets are used to refine the current partition.

In our original version of the~PR reordering algorithm~\cite{JNP18},
partitions for all of the supernodes are being refined simultaneously
throughout the course of the computation.
The reader should consult~\cite{JNP18} for a detailed description of this 
algorithm; it is detailed enough that it could probably be used
to code up the algorithm.
We have replaced this implementation with one that processes and
completely refines a partition for each supernode, one at a time.
Our description of the new implementation
is a high-level description,
and it uses a great deal of notation and terminology that
has not been introduced yet in the paper.
Consequently, we will proceed as follows.
We will walk the reader from beginning to end line-by-line
through the algorithm,
introducing and describing all notation and terminology
as it is encountered during the process.
The algorithm can be found in Algorithm~\ref{alg:PR}.

\begin{algorithm}[htp] % :set ts=2
%\footnotesize
%\small
\begin{flushleft}
	{\bf Input:}
	The supernodes $J_1,J_2,\ldots,J_N$; \\
	the higher adjacency sets $\hadj{G^{+}}{J_s}$ for $s = 1,2,\ldots,N$; \\
	{\bf Output:}
	A reordering of the vertices of each supernode consistent with \\ 
	the final ordered partition of the supernode.
\end{flushleft}
\caption{Our high-level~PR reordering algorithm that processes one supernode at a time.}
\label{alg:PR}
	\begin{algorithmic}[1]
	\For{$t \gets 1$ {\bf to} $N$}
		\State{$S \gets J_t$; ${\mathcal P} \gets [S]$;}
		\For{each supernode $J_s \in \Tree{J_t}$ in a reverse topological ordering}
			\State{Use $\hadj{G^{+}}{J_s} \cap J_t$ to partition the ``partitionable'' sets in~${\mathcal P}$;}
			\LineComment{Refine the partition \ldots}
			\For{each ``partitionable interval''~$I$ in~${\mathcal P}$}
				\State{$S' \gets \emptyset$;}
				\For{each set $S \in I$ (in order)}
				\LineComment{\ldots alternating between $S \cap \hadj{G^+}{J_s}$ to the right and $S \cap \hadj{G^+}{J_s}$ to the left}
					\If{$S$ is first in $I$ or $S$ is preceeded in ${\mathcal P}$ by the ordered pair $S' \cap \hadj{G^+}{J_s}$, $S' \setminus \hadj{G^+}{J_s}$}
						\State{Replace $S$ in ${\mathcal P}$ with the pair $S \setminus \hadj{G^+}{J_s}$, $S \cap \hadj{G^+}{J_s}$, in that order;}
					\Else
						\State{Replace $S$ in ${\mathcal P}$ with the pair $S \cap \hadj{G^+}{J_s}$, $S \setminus \hadj{G^+}{J_s}$, in that order;}
					\EndIf
					\State{$S' \gets S$;}
				\EndFor
			\EndFor
		\EndFor
		\State{Reorder supernode $J_t$ consistent with the ordered partition in ${\mathcal P}$;}
	\EndFor
\end{algorithmic}
\end{algorithm}

The input consists of the supernodes $J_1,J_2,\ldots,J_N$ and the so-called
higher adjacency sets of the supernodes.
To formally define these sets requires the use of the fill graph~$G^+$ of the 
factor matrix.
For our purposes, it suffices to point out that,
in Figure~\ref{fig:supernode1},
we have $\hadj{G^+}{J_1} = \{5, 6, 9\}$,
$\hadj{G^+}{J_2} = \{5, 7, 8\}$, and
$\hadj{G^+}{J_3} = \emptyset$, and,
in Figure~\ref{fig:supernode2},
we have $\hadj{G^+}{J_1} = \{5, 6, 7\}$,
$\hadj{G^+}{J_2} = \{7, 8, 9\}$, and
$\hadj{G^+}{J_3} = \emptyset$.

The output for each supernode is an ordering of the supernode
consistent with the final ordered partition of the supernode.
Each iteration of the outer {\bf for} loop beginning at line~1
and ending at line~18 processes a single supernode~$J_t$
from among the~$N$ supernodes.
Line~2 initializes set~$S$ to be~$J_t$, and it initializes the
ordered partition~${\mathcal P}$ so that it contains the
single set~$S$.

Each iteration of the {\bf for} loop beginning at line~3
and ending at line~16 processes a supernode~$J_s$ whose set
$\hadj{G^+}{J_s} \cap J_t$ is needed to 
partition sets in~${\mathcal P}$ and ultimately
to complete a refining step of the partition.
The set~$\Tree{J_t}$ (in line~3) from which the supernodes~$J_s$ are taken
is a subtree of the {\em supernodal elimination tree} 
consisting of every supernode~$J_s$ that updates supernode~$J_t$.
(For more on the relevant properties of elimination trees,
consult Liu~\cite{Liu90}.)
A {\em topological ordering} of a rooted tree
numbers every child before its parent;
a {\em reverse topological ordering} (in line~3) numbers every parent
before each of its children.

Let~$S$ be a set in~${\mathcal P}$.
In line~4, set~$S$ is deemed a ``partitionable'' set if 
\begin{displaymath}
  S \cap \hadj{G^+}{J_s} \cap J_t = S \cap \hadj{G^+}{J_s} \neq \emptyset
\end{displaymath}
and
\begin{displaymath}
  S \setminus \hadj{G^+}{J_s} \cap J_t = S \setminus \hadj{G^+}{J_s} \neq \emptyset.
\end{displaymath}
In line~4, each of these ``partitionable'' sets is partitioned into a pair of sets,
as described above.

The details of completing the refinement of~${\mathcal P}$ by appropriately
incorporating the pairs of sets found in line~4 occur in the {\bf for}
loop beginning in line~5 and ending in line~15.
An alternating strategy, first used in Jacquelin, et al.~\cite{JNP18},
is also used in this loop.

In line~5, a ``partitionable interval'' in~${\mathcal P}$ simply refers to the following.
A set of {\em consecutive} sets~$I$ in~${\mathcal P}$ is a ``partitionable interval''
if
\begin{enumerate}
  \item
    Every set $S \in I$ is ``partitionable''.
  \item
    No ``partitionable'' set $S' \in {\mathcal P} \setminus I$ can be
    added to~$I$ to create a longer ``partitionable interval''.
\end{enumerate}
The {\bf for} loop starting in line~7 and ending in line~14
processes the sets in a ``partitionable interval'' in order from
left to right, so to speak.
The reader may easily verify that during the odd numbered iterations
through this loop, the set $S \cap \hadj{G^+}{J_s}$ becomes second
(rightmost) of the pair of sets, and
during the even numbered iterations
through this loop, the set $S \cap \hadj{G^+}{J_s}$ becomes first
(leftmost) of the pair of sets.
Alternating in this fashion roughly cuts in half the number
of $(J_t,J_s)$-blocks produced, compared to the strategy of
always moving $S \cap \hadj{G^+}{J_s}$ to the left or always
moving it to the right.

As mentioned earlier, the implementation of~PR reordering in Algorithm~\ref{alg:PR}
greatly reduces the time and working storage requirements
over the implementation introduced in Jacquelin, et al.~\cite{JNP18}.
The initial implementation required many $n$-vectors of working storage.
Let $m' = \max_{1 \leq t \leq N} |J_t|$.
Our implementation of Algorithm~\ref{alg:PR} requires only one $n$-vector,
along with seven $N$-vectors and fourteen $m'$-vectors.
Because it is generally the case that both~$N$ and~$m'$ are very small relative
to~$n$,
the working storage requirements for Algorithm~\ref{alg:PR} are reduced to
ideal levels, in practice.

We conjecture that working on one supernode at a time, using a data structure
that occupies much less space, greatly reduces the number of cache misses
during the computation.
We conjecture that this probably explains the large reductions in runtimes
over those obtained using the original implementation in~\cite{JNP18}.
We choose not to compare runtimes for Algorithm~\ref{alg:PR} with those of the
earlier implementation of~PR reordering.
We will wait until Section~\ref{sec:TSP-PR},
where we will compare (indirectly) the runtimes for Algorithm~\ref{alg:PR}
against those of~TSP reordering.

We close this subsection by presenting a method to modestly improve~PR reorderings.
Jacquelin, et al.~\cite{JNP18} reported on two specific 
{\em reverse topological orderings} (see line~3 of Algorithm~\ref{alg:PR})
that improved~PR reordering quality over the ``natural'' ordering.
The two orderings performed very similarly to one another in Jacquelin, et al.~\cite{JNP18}.
Here, we will focus on just one of these orderings:
the {\em maximum descendants} reverse topological ordering.
For this ordering, the next supernode to be selected from among the eligible supernodes
is that with the maximum number of descendants in the supernodal
elimination tree.

\begin{figure}[htb]
\begin{center}
  \caption{Performance profile for RLB factorization times using two different versions
           of~PR reorderings.}
  \includegraphics[width=.8\textwidth]{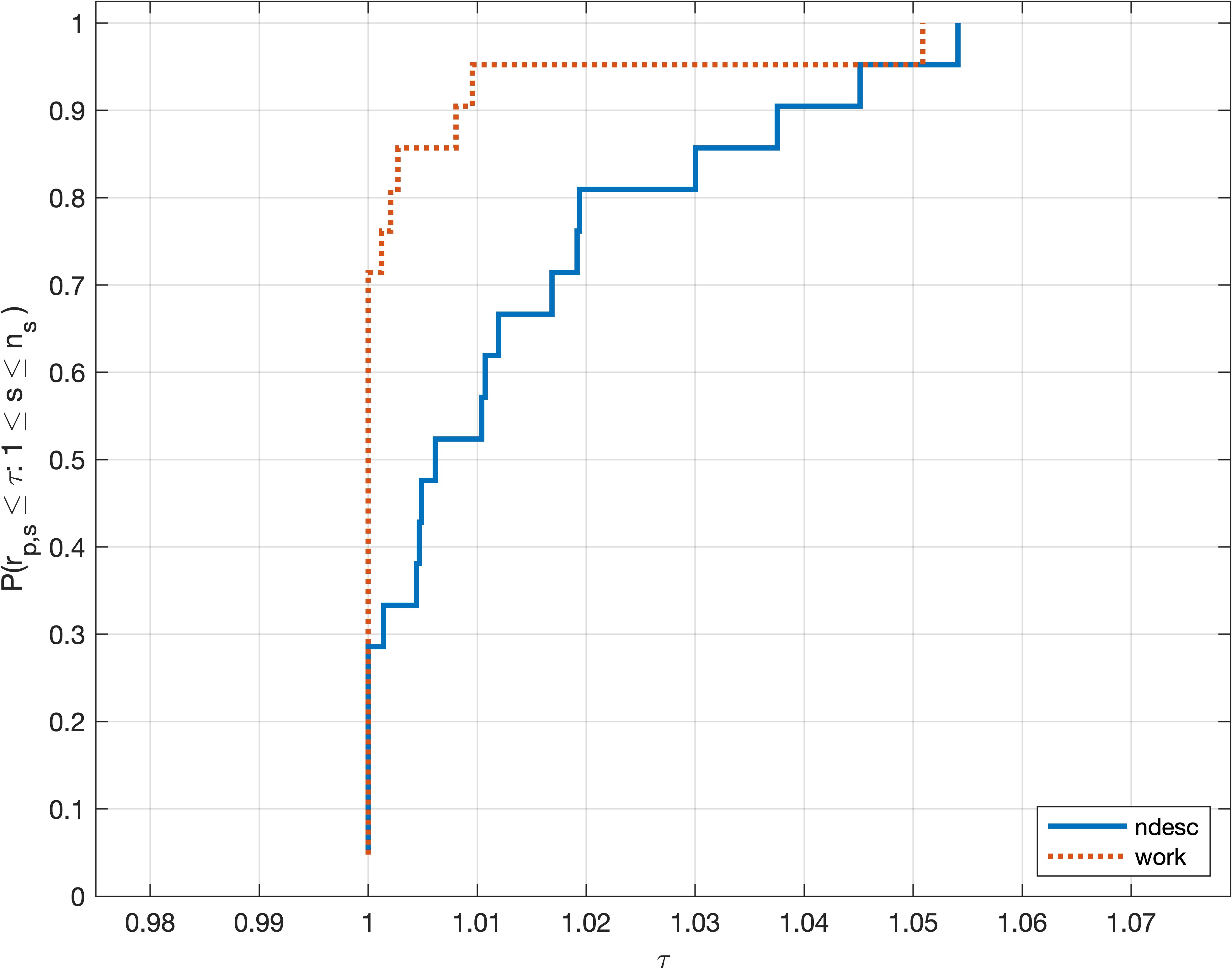}
  \label{fig:PR}
\end{center}
\end{figure}

We have, however, found that the following strategy works slightly better:
Choose from among the eligible supernodes that supernode~$J'$
whose subtree (rooted at~$J'$) in the supernodal elimination tree requires the
{\em most} work to factor.
In Figure~\ref{fig:PR}, the reader will find a performance profile
for~RLB factorization times using the two different reverse topological orderings
discussed above in Algorithm~\ref{alg:PR} prior to each factorization.
The one involving the number of descendants is identified by ``ndesc'',
and the one involving factorization work is identified by ``work''.

Clearly, using the factorization work in this fashion produces
modestly, but unambiguously,  better results.
The option based on work is fastest for approximately
seventy percent of the matrices.
Moreover, for all but one of the other matrices,
the option based on work requires no more than a one percent
penalty in the runtime.

%
% partitioning into paths for Suporder.tex
%

\subsection{A comparison of the best TSP and PR reorderings}
\label{sec:TSP-PR}

In Section~\ref{sec:TSP}, we improved our implementation of the~TSP
reordering method, and,
in Section~\ref{sec:PR}, we improved our implementation of the~PR reordering method.
Henceforth,~TSP will refer to the best version of~TSP presented in Section~\ref{sec:TSP}:
one using farthest insertion and weights in the objective function.
Henceforth,~PR will refer to the best version of~PR presented in Section~\ref{sec:PR}:
one based on Algorithm~\ref{alg:PR} that uses the work-based
criterion to determine the reverse topological orderings.

This subsection contains the key results presented in this paper:
a clear and fair comparison of the~TSP and~PR methods.

\begin{figure}[htb]
\begin{center}
  \caption{Performance profile for RLB factorization times, 
           with and without the reordering overhead, using
           the best~TSP and~PR reorderings.}
  \includegraphics[width=.8\textwidth]{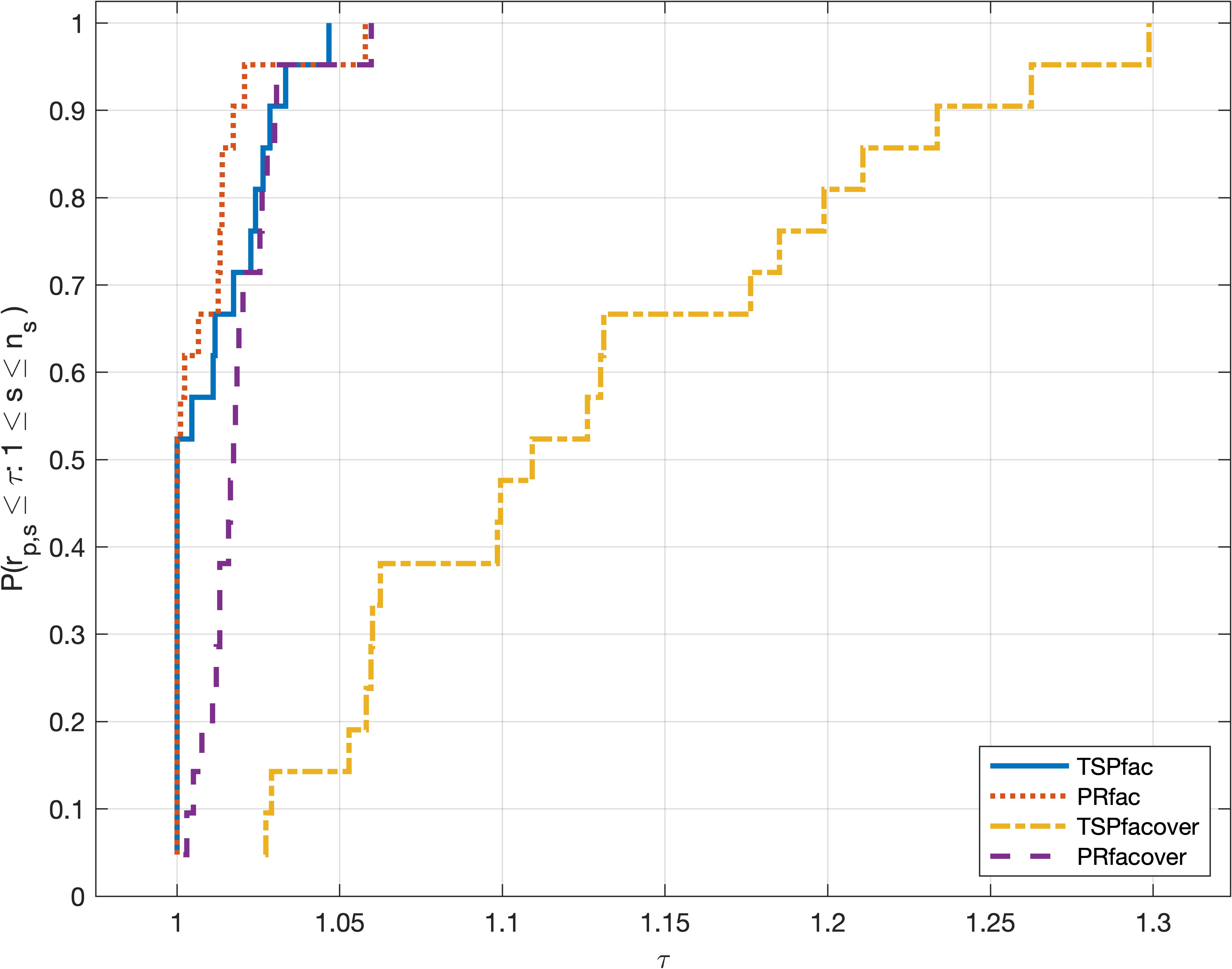}
  \label{fig:time}
\end{center}
\end{figure}

To perform a fair comparison of how effective the two methods are at
reducing factorization times, we present the factorization times both
with and without the overhead time required to perform the reordering of the 
supernodes.
(As a technical note, the overhead time also includes the time required by an additional symbolic
factorization that must be performed.)
The results of this experiment are recorded in the performance profile in Figure~\ref{fig:time}.
The lines marked ``TSPfac'' and ``PRfac'' are for the~TSP and~PR
factorization times (respectively) without accounting for any~TSP or~PR overhead time.
Observe that the lines track each other very closely,
but if one had to give the advantage to one method over the other,
it would probably be~PR.

The lines marked ``TSPfacover'' and ``PRfacover'' provide a much fairer comparison.
The data for these lines include the overhead time required to reorder the supernodes.
Clearly, the~TSP reordering times are very large relative to the~PR
reordering times.
Our conclusion from this chart is that~PR should be the method of choice.

\begin{figure}[htb]
\begin{center}
  \caption{Performance profile for the work space requirements of the best~TSP and~PR reorderings.}
  \includegraphics[width=.8\textwidth]{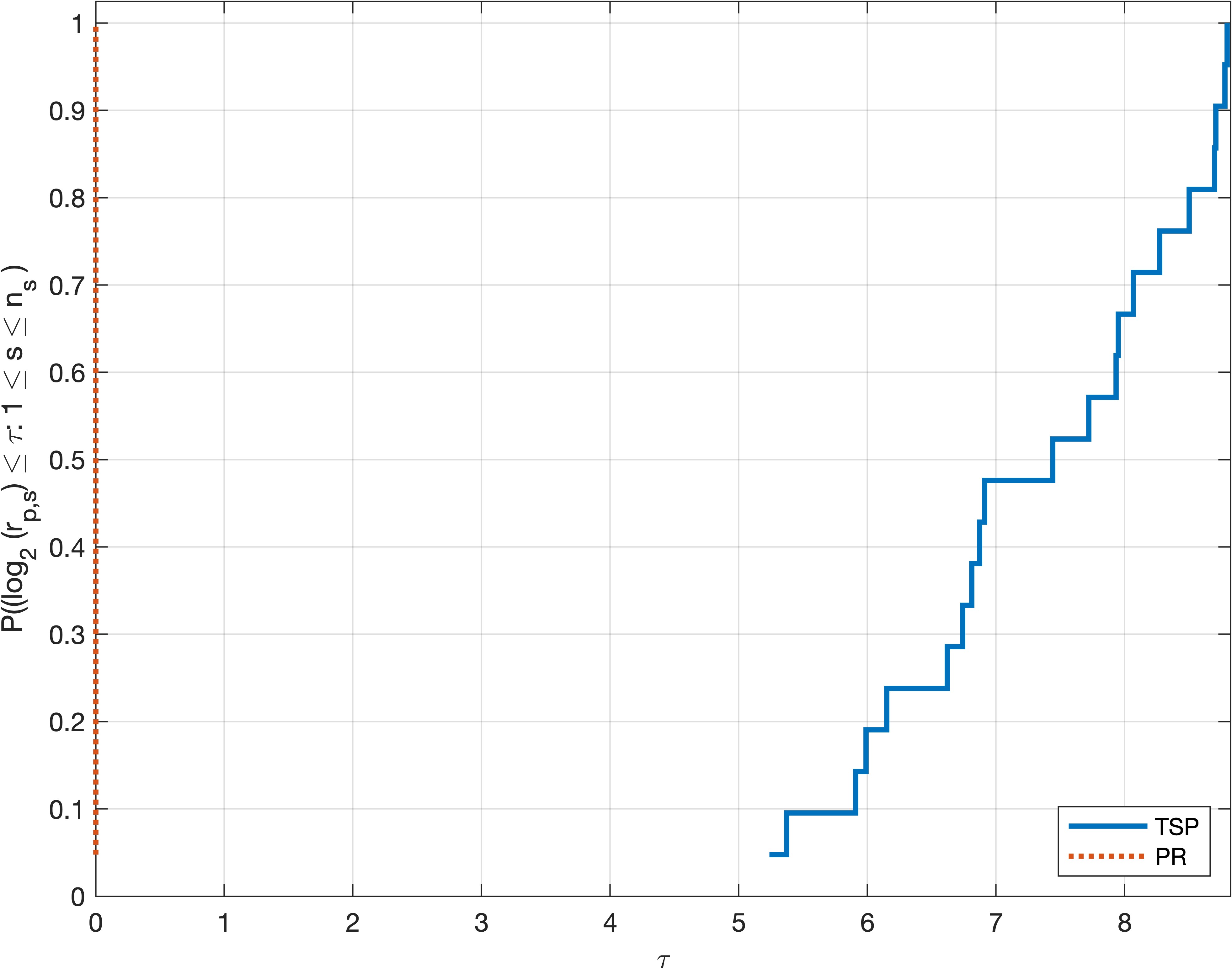}
  \label{fig:space}
\end{center}
\end{figure}

This conclusion is reinforced when we look at the
working space requirements for the two methods.
The performance profile in Figure~\ref{fig:space}
compares the working storage requirements for both methods,~TSP and~PR.
Because the working storage requirements for~TSP routinely exceed that required by~PR
by orders of magnitude, the data for the two methods are compared logarithmically.
Clearly, our conclusion is that~PR should be the method of choice by this criterion
also.

%
% partitioning into paths for Suporder.tex
%

\section{Conclusion}
\label{sec:conclusion}

In this paper, we made every effort to perform a fair and meaningful
comparison of the~TSP and~PR reordering methods.
We compared them primarily by measuring their impact on RLB factorization
times on~48 cores of our test machine.
RLB is particularly well suited for this purpose because it is entirely dependent
for good performance on reorderings of the supernodes that lower the number
of blocks and increase the block sizes.

In previous papers~\cite{JNP18,JNP21,PFRR17},
researchers took significant steps toward turning the~TSP and~PR reordering algorithms
into mature and well-developed algorithms, with well-designed and efficient
implementations.
In Sections~\ref{sec:TSP} and~\ref{sec:PR},
we have taken further steps that improve both algorithms and their
implementations.
Particularly important in this regard is Algorithm~\ref{alg:PR}
in Section~\ref{sec:PR},
which improves the implementation of~PR immensely.

Essentially, our best versions of~TSP and~PR are roughly equally effective
at bringing~RLB factorization times down, but~PR is the clear choice over~TSP
because the time and working storage required to compute~PR reorderings is so 
much less than that required to compute~TSP reorderings.
We conjecture that this will remain the case in the future;
it will be interesting to see if these two methods are subject to significant
improvements in the future.

\section*{ACKNOWLEDGMENTS}
This work was supported in part by the U.S. Department of Energy, Office of Science, Office of Advanced Scientific
Computing Research and Office of Basic Energy Sciences, Scientific Discovery through Advanced Computing (SciDAC)
Program through the FASTMath Institute and BES Partnership under Contract No. DE-AC02-05CH11231 at Lawrence
Berkeley National Laboratory.

\bibliographystyle{ACM-Reference-Format}
\bibliography{mlf_paper}

\end{document}